\documentclass[a4paper,11pt]{article}

\usepackage{amsmath}
\usepackage{amssymb}
\usepackage{graphicx, color}
\usepackage{palatino}
\usepackage{overcite}
\newcommand{\sups}[1]{$^{#1}$}

\makeatletter

\def\@biblabel#1{#1.}

\makeatother

\begin{document}

\begin{titlepage}

\vspace{20 pt}

\noindent\textsf{\textbf{\Large A generic mechanism of emergence of amyloid protofilaments from disordered oligomeric aggregates}}

\vspace{20 pt}

\noindent\textsf{\textbf{Stefan Auer\sups{1,\ast}, Filip Meersman\sups{2}, Christopher M. Dobson\sups{3}, and Michele Vendruscolo\sups{3,\ast}}}

\date{\today}

\pagenumbering{arabic}

\vspace{80 pt}

\noindent\sups{1}\textsl{Centre for Self Organising Molecular Systems, 
University of Leeds, Leeds LS2 9JT, UK}

\vspace{10 pt}

\noindent\sups{2}\textsl{Department of Chemistry, Katholieke Universiteit Leuven, Celestijnenlaan 200F, B-3001 Leuven, Belgium}

\vspace{10 pt}

\noindent\sups{3}\textsl{Department of Chemistry, Lensfield Road, Cambridge,
CB2 1EW, UK} \\

\vspace{10 pt}
\noindent\sups{\ast} \textsl{Corresponding Authors
(\texttt{s.auer@leeds.ac.uk,mv245@cam.ac.uk}).}

\vspace{60 pt}

\vspace{20 pt}

\end{titlepage}

\section*{Abstract}
The presence of oligomeric aggregates, which is often observed
during the process of amyloid formation, has recently attracted much
attention since it has been associated with neurodegenerative conditions such as Alzheimer's and Parkinson's diseases. 
We provide a description of a sequence-indepedent mechanism by which polypeptide chains aggregate by forming metastable oligomeric 
intermediate states prior to converting into fibrillar structures. 
Our results illustrate how the formation 
of ordered arrays of hydrogen bonds drives the formation of $\beta$-sheets within the 
disordered oligomeric aggregates that form early under the effect of hydrophobic forces. 
Initially individual $\beta$-sheets form with random orientations, which 
subsequently tend to align into protofilaments as their lengths increases. 
Our results suggest that amyloid aggregation represents an example of
the Ostwald step rule of first order phase transitions by showing that
ordered cross-$\beta$ structures emerge preferentially from disordered
compact dynamical intermediate assemblies.
\vspace{10truemm}

\section*{Author Summary}

Considerable efforts are currently devoted to the study of the
phenomenon of protein aggregation because of its association with a
range of human diseases and of its potential applications in
biotechnology.  Despite intense scrutiny, however, it has been
extremely challenging so far to characterise the intermediate phases
of the aggregation process, during which disordered oligomeric
assemblies are formed.  In our work, we have used molecular
simulations to show that the development of ordered structures within
the initially disordered aggregates is a consequence of the interplay
between hydrophobicity and hydrogen bonding, which results in a
behaviour typical of many first order 
condensation phenomena in material science.
These results provide further insight into the consequences of the
``generic hypothesis" of protein aggregation, according to which the
ability to assemble into ordered cross-$\beta$ structures is not an
unusual feature exhibited by a small group of peptides and proteins
with special sequence or structural properties, but it is an inherent
characteristic of polypeptide chains.

\clearpage

\section*{Introduction}

A variety of peptides and proteins unrelated in sequence and structure
have been shown to convert into large ordered aggregates known as
amyloid fibrils \cite{chiti06,jahn08}.  These structures share a common
cross-$\beta$ structure formed by intertwined layers of $\beta$-sheets
extending in a direction parallel to the fibril axis \cite{makin05,chiti06}.
The ubiquity of this type of assemblies has led to the suggestion that
they may represent a general structural state of polypeptide chains,
which is accessible independently from their specific amino acid
sequences \cite{dobson99a}.  According to this view, if placed under
appropriate conditions, peptides and proteins can revert to the
amyloid state, which has been associated with a range of
pathological conditions including Alzheimer's and Parkinson's diseases
\cite{selkoe03,stefani03,chiti06}.

Small oligomeric aggregates are often found as precursors of amyloid
fibrils \cite{harper97a,lambert98,serio00}, and their formation in
some cases may originate from a competition between amorphous and
fibrillar aggregation.  The role of these molecular species in the
process of amyloid fibril formation is at present unclear, although
much interest has been recently devoted to this problem since their
presence has been linked to neurodegenerative processes
\cite{lambert98,haass07}.  It has been suggested that, under
conditions that favor amyloid fibril formation, proteins or peptides
within these disordered aggregates can convert into conformations
capable of forming nuclei that give rise to amyloid fibril assemblies
\cite{serio00}.  It has been, however, extremely challenging to characterize
experimentally the structures of these aggregates and the mechanism of
their formation owing to their heterogeneous and dynamical nature.

In this work we consider the problem from a theoretical perspective
and use computer simulations to describe the process of condensation
of polypeptide chains into oligomeric assemblies that further
reorganise into fibrillar structures. The level of detail in which
protein aggregation can be investigated depends on the choice of the
model.  Full-atomistic simulations have provided considerable insight
into the dynamics of inter-molecular interactions in systems
containing a small number of peptides and short timescales
\cite{ma02,hwang04a,buchete05,hills07,nguyen07,cheon07}.
Complementary to these approaches, coarse-grained models have proven
capable of simulating larger systems and longer timescales, and of
following the structure of the oligomeric intermediates and the
mechanism of their conversion into ordered cross-$\beta$ assemblies
\cite{nguyen04,urbanc04,pellarin06,pellarin07,derreumaux07}. Despite much recent
work in this area, many questions about the amyloid aggregation remain
open, and here we investigate the general properties of the mechanism
of emergence and alignment of $\beta$-sheets in the early stages of
the oligomerization process. Given the close link between this phase
of amyloid formation and the neurotoxicity of the structural species
involved \cite{lambert98,chiti06,haass07,luheshi07}, we investigated
here the competition between ordered and disordered aggregation of
polypeptide chains.

By following the hypothesis that amyloid formation represents a
generic property of a polypeptide chain \cite{dobson99a}, we adopt a
recently proposed representation of the structure of polypeptide chains, known as the tube model \cite{hoang04,hoang06,auer07a}, 
which enables a description of the free energy landscapes
for folding \cite{hoang04,hoang06,auer07a} 
and for aggregation \cite{auer07} to be obtained within a unified
framework. Since this model only includes
interactions common to all polypeptide chains independently
from their amino acid sequence, it is ideally
suited for exploring the consequences of the generic hypothesis of
amyloid formation.  The characteristic features of the model
\cite{hoang04} are that the protein backbone is assigned a finite
thickness to account for excluded volume effects. Residues interact
with each other by pairwise additive hydrophobic forces (with energy
$e_W$), geometrical constraints apply to the formation of intra- and
intermolecular hydrogen bonds (with energy $e_{HB}$), and the
polypeptide chain experiences a local bending stiffness (with energy
$e_S$). 

\section*{Results}

\begin{figure*}
\begin{center}
\includegraphics[width=180mm]{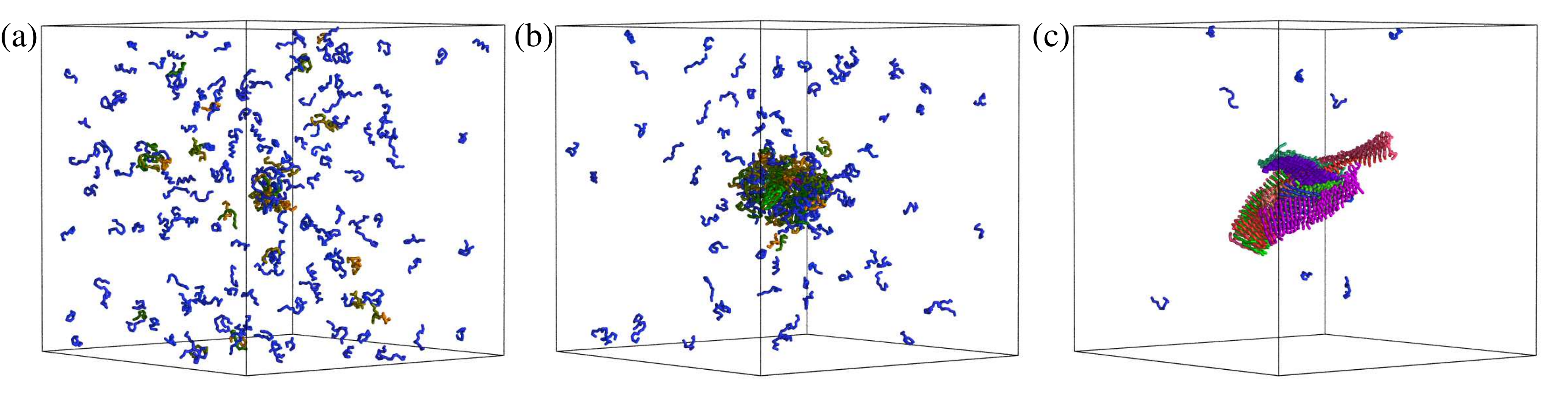}
\caption{
\label{fig:box}
Illustration of the self-assembly process of peptides into amyloid-like
assemblies. All simulations were carried out at a concentration $c=12.5$ mM
and reduced temperature $T^*=0.66$. 
The progress variable $t$ corresponds to the number of Monte Carlo
moves performed in the simulation, and one unit of $t$ is a series of $10^5$
Monte Carlo moves.  Initially, at $t=1000$ (left panel), all peptides are in a
solvated state.  As the simulation progresses ($t = 5000$, middle panel) a
hydrophobic collapse causes the formation of a disordered oligomer, which
subsequently undergoes a structural reorganization into an amyloid-like
assembly ($t = 30000$, right panel) driven by the formation of ordered arrays
of hydrogen bonds.  Peptides that do not form intermolecular hydrogen bonds
are shown in blue, while peptides that form intermolecular hydrogen bonds are
assigned a random color, which is the same for peptides that belong to same
$\beta$-sheet.}
\end{center}
\end{figure*}

In this work we consider a system containing $216$ $12$-residue
homopolymers that exibit an $\alpha$-helical native state below the
folding temperature ($T^*_f\sim 0.58$) and an undfolded structure at
higher temperatures (see Methods). Peptides that form native
$\alpha$-helical conformations \cite{kammerer04}, as well as
homopolymeric sequences \cite{fandrich02}, have been shown to be able
to form amyloid assemblies. In order to investigate the self-assembly
of the peptides into fibrils we chose thermodynamic conditions
such that fibril formation occurs on a timescale accessible to our
simulations.  We found that a peptide concentration $c=12.5mM$ is
above the critical concentration for aggregation, provided that the
temperature $T^*<0.69$. All our simulations were performed at
$T^*=0.66$, and several independent starting configurations were generated
at $T^*=0.75$. As in our simulations we set $T^* > T^*_f$, 
the peptides were unfolded most of the time. 
A typical trajectory observed in
our Monte Carlo simulations (see Methods) is illustrated in
Fig.~\ref{fig:box}. 

\begin{figure}
\begin{center}
\includegraphics[width=85mm]{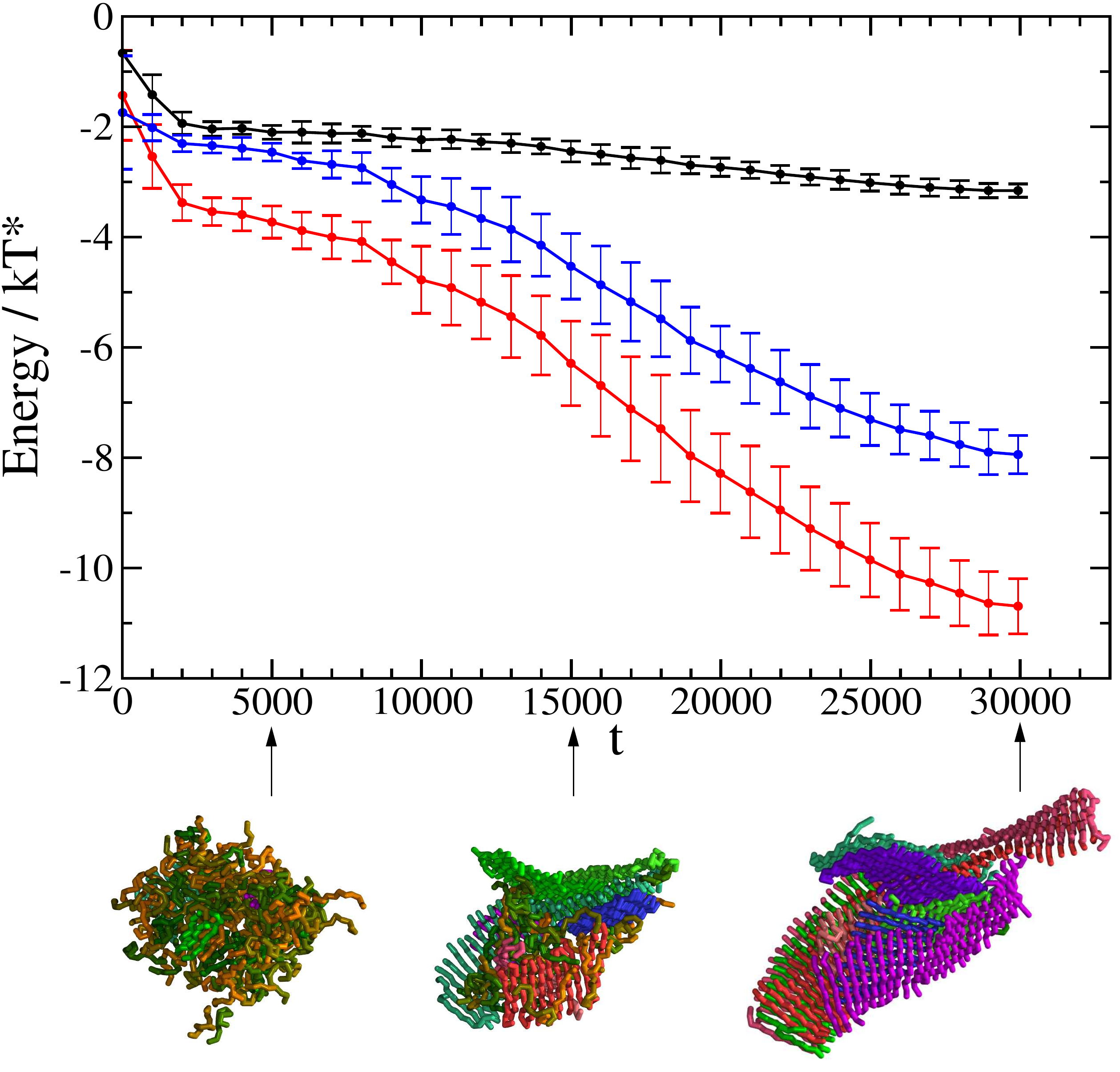}
\caption{
\label{fig:growth}
Total energy per peptide as a function of $t$.  
Structures formed during the process of conversion of the disordered oligomer
into an amyloid-like structure are also shown at
$t=5000$, $t=15000$, and $t=30000$. The color
code is as in Fig. \ref{fig:box}.}
\end{center}
\end{figure}

We systematically observed a rapid collapse of the peptides into
disordered aggregates that subsequently underwent a structural
reorganization and transform into cross-$\beta$ protofilaments (Fig.
\ref{fig:box}). These results are consistent with a
previously described two-step condensation-ordering mechanism
\cite{nguyen04,auer07,cheon07}, which has also been observed
experimentally \cite{serio00}. A plot of the total energy per peptide
as a function of the progress variable $t$ (Fig.  \ref{fig:growth})
shows that the final structure has a much lower energy than the
initial and intermediate states. The major contribution to this
energy comes from hydrogen bonding (Fig. \ref{fig:growth}), 
a result consistent with
the recent report that the hydrogen bonding energy provides
the dominant factor stabilising the cross-$\beta$ architecture
is represented by hydrogen bonding, while in more disordered states
other contributions are also important \cite{knowles07s}.
The initial state ($t<1000$), before the hydrophobic collapse, in
which all peptides are solvated, has the highest energy and it is
unstable. After the hydrophobic collapse has taken place
($1000<t<5000$), the
peptides form a disordered oligomer, which is characterised by
similar contributions from hydrophobic interactions and hydrogen
bonding (Fig. \ref{fig:growth}); this oligomeric state is lower in energy but metastable
with respect to the amyloid state.  Finally, with the growth of
the cross-$\beta$ architecture the hydrogen bonding
interactions become progressively dominant (Fig. \ref{fig:growth}). 
The survival time of the
disordered oligomeric state is rather short (about 10-15\% of the total
simulation time) since in order to be able to investigate the
self-assembly of the peptides we chose thermodynamic conditions such
that the nucleation barriers associated with oligomer formation and
the subsequent ordering are readily overcome by thermal fluctuations.
The height of the nucleation barriers, and the associated lag times
depend strongly on the thermodynamic conditions of the system
\cite{auer07}.

In order to provide a detailed description of the emergence of
cross-$\beta$ protofilaments within the oligomers, including their
interactions and relative orientations with respect to each other, we
defined the oligomeric state using a distance criterion that requires
the centres of mass of two peptides to have a distance of less than
$5\AA$. Two peptide chains are taken to form a $\beta$-sheet if they
have more than four inter-chain hydrogen bonds with each other. To
define an angle between different $\beta$-sheets we calculated the
relative orientation between neighboring peptides that constitute the
different $\beta$-sheet. Therefore we calculate the dot product of the
end to end vectors of the peptide molecules, requiring that the
centres of mass of two peptides are separated by less than $10\AA$,
which is the typical inter-sheet contact distance in most native and
amyloid systems. If the average angle between two $\beta$-strands is
less than $20$ degree, we assume that the respective $\beta$-sheets
belong to the same protofilament.

\begin{figure}
\begin{center}
\includegraphics[width=85mm]{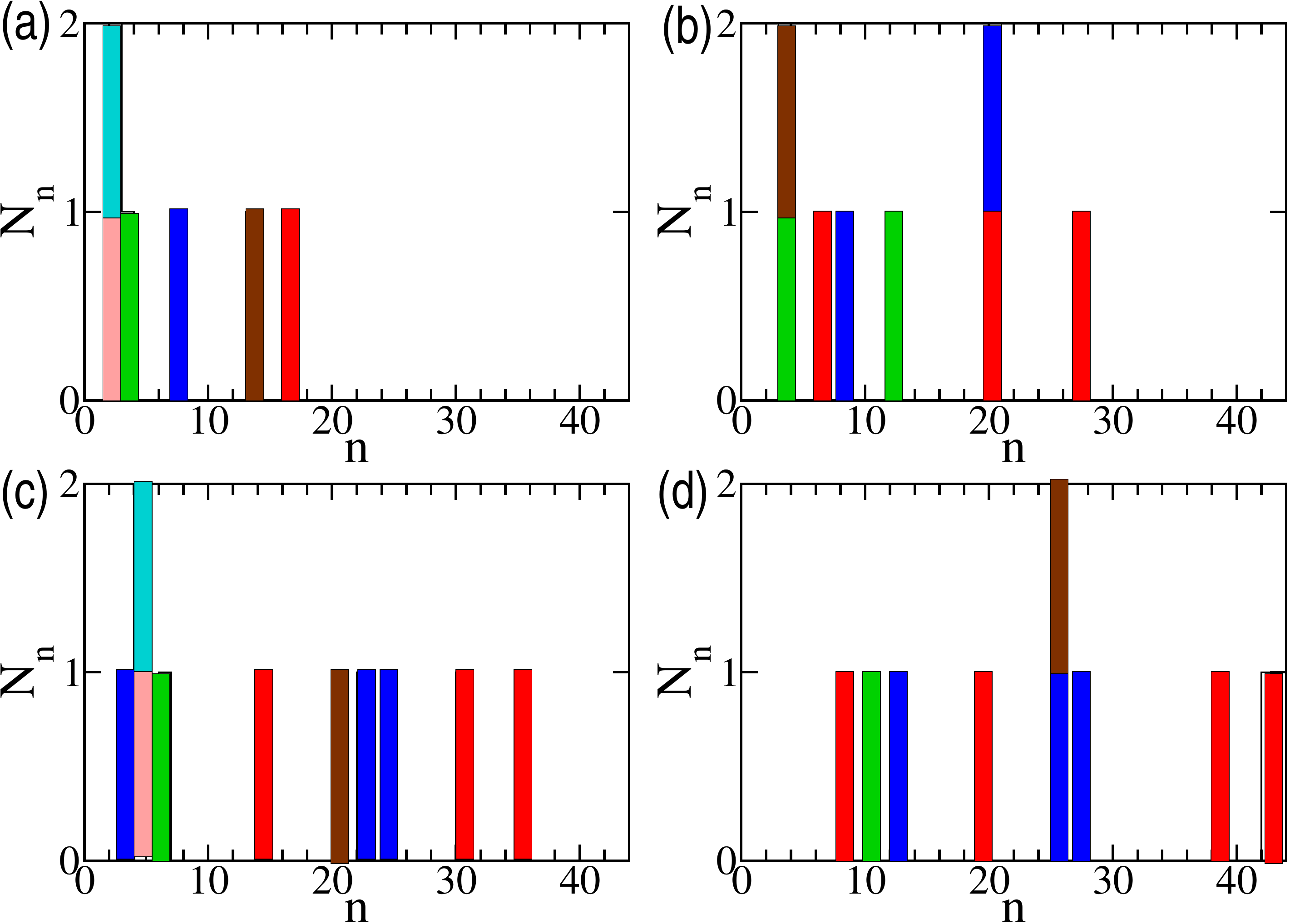}
\caption{
\label{fig:Hsheets}
Histogram of the number $N_n$ of $\beta$-sheets consisting of $n$
peptides at four successive stages of the growth and reordering
process of the oligomeric assembly shown in Fig. \ref{fig:box}: (a)
$t=10000$, (b) $t=15000$, (c) $t=20000$, (d) $t=30000$). This plot
shows how $\beta$-sheet assemblies are progressively formed by the
growth and alignment of individual $\beta$-sheets.  At $t=10000$ (a)
there are six $\beta$-sheets of sizes ranging from $3$ to $16$,
whereas at $t=30000$ (d), there are nine $\beta$-sheets of sizes
ranging from $8$ to $42$. If $\beta$-sheets are aligned so that the
angle between them is smaller than $20$ degrees, they are considered
to form a protofilament-like structure, and the corresponding bars in
the histogram are shown with the same color, as for instance in the
case of the red assembly (Fig. \ref{fig:box}c right panel), formed by
four $\beta$-sheets of size $8$, $19$, $38$ and $42$.}
\end{center}
\end{figure}

In the example illustrated in Fig. \ref{fig:box}, the initial stages
of the process are characterized by the formation within the
disordered oligomer of six small $\beta$-sheets which are randomly
oriented with respect to each other (Fig. \ref{fig:Hsheets}a).
Subsequently, the $\beta$-sheets tend to align as their lengths
increase, and protofilaments consisting of one, three and four
$\beta$-sheets are formed (Fig.  \ref{fig:Hsheets}b-d). 
The two major protofilaments observed in this simulation
seem to twist around each other (Fig.  \ref{fig:box} right panel),
resembling the typical behavior observed experimentally
\cite{chiti06}. The twisting appears to follow from the growth and
alignment of $\beta$-sheets, which is a consequence of the tendency to
optimize the number of hydrophobic contacts, thereby reducing the
interfacial energy \cite{turner03}, and not from the chirality of the
peptides, as the latter is not included in the tube model used in this
work. As the peptides within the oligomer can move only locally our
Monte Carlo dynamics should at least qualitatively resemble their actual
dynamics.

\begin{figure}
\begin{center}
\includegraphics[width=85mm]{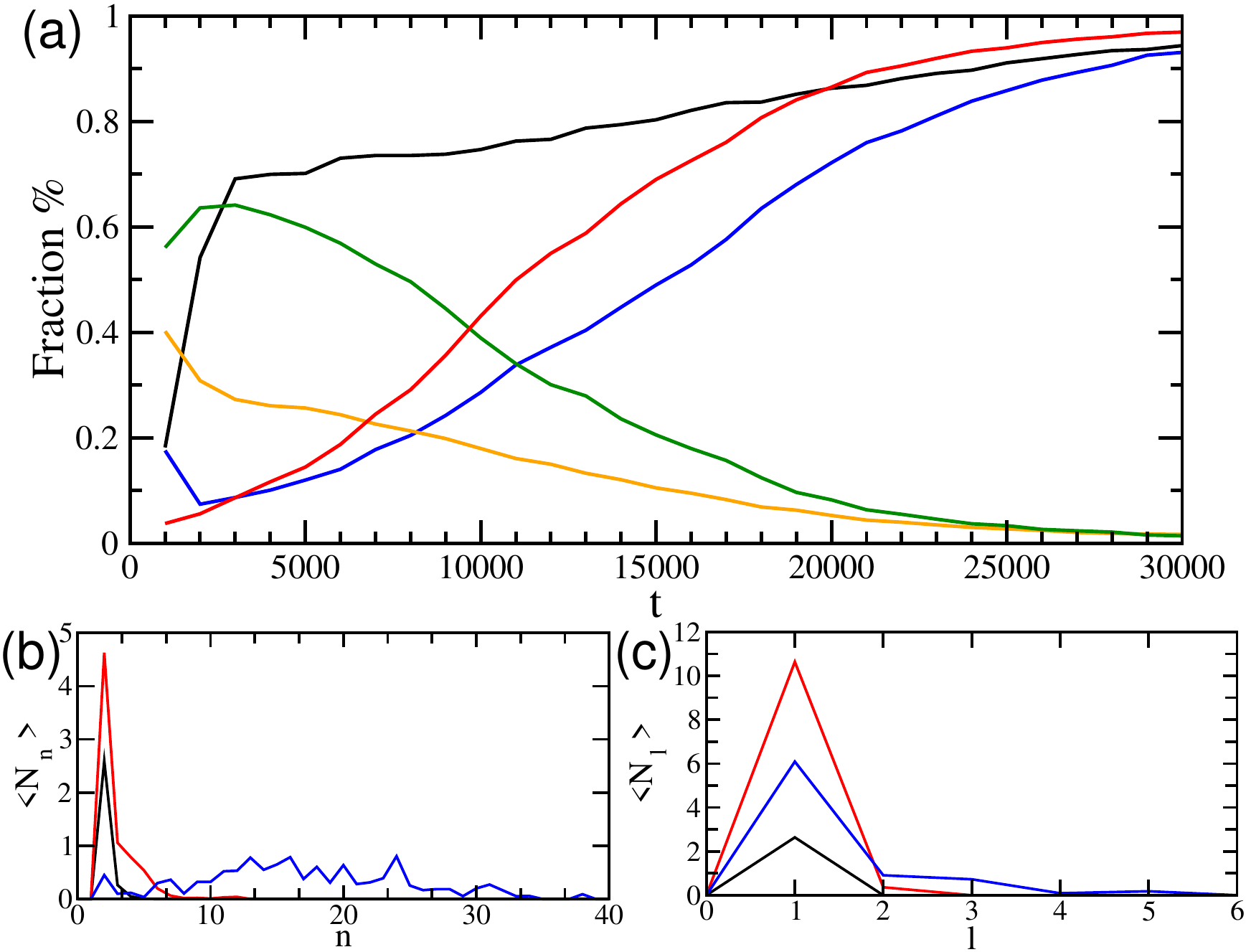}
\caption{
\label{fig:11runs}
Analysis of the evolution of the structure of the oligomers over $11$
independent simulations.  (a) Development of the fraction of polypeptide
chains in a oligomer (black), fraction of polypeptide chains in a oligomer
that form a $\beta$-sheet conformation (blue), fraction of hydrogen bonds in a
oligomer in a $\alpha$-helical conformation (orange), and in a $\beta$-sheet
conformation (red), or otherwise (green). (b) Development of the distribution
function of the average number of $\beta$--sheets $\langle N_n \rangle$ of
size $n$ at $t=1000$ (black), $t=5000$ (red), $t=30000$ (blue). (c)
Distribution function $\langle N_l \rangle$ of the number of protofilaments
composed of $l$ layers at $t=1000$(black), $t=15000$(red), $t=30000$(blue).  }
\end{center}
\end{figure}

We generated and analyzed a total of $11$ independent trajectories,
which consistently appeared as the type shown in Fig. \ref{fig:box},
and showed the same quantitative overall behavior. Assemblies are
initially formed through the disordered rapid assembly of partially
folded peptides, which then reorganize into ordered $\beta$ sheets.  A
quantitative analysis (Fig. \ref{fig:11runs}) of the reordering
process shows that initially about $60\%$ of the hydrogen bonds within
the oligomers are formed in disordered intermolecular associations,
whereas the remainder are involved in intramolecular interactions
within the native $\alpha$-helix conformation (Fig.
\ref{fig:11runs}a).  At later stages, a structural reorganization of
the oligomers results in essentially all hydrogen bonds being involved
in the cross-$\beta$ structure. Thus, in agreement with experimental
evidence \cite{dirix05,petty05,knowles07}, we found that the formation
of disordered oligomers is primarily driven by hydrophobic effects,
whereas a reorganisation driven by hydrogen bond formation is
subsequently playing a major role in the formation of cross-$\beta$
structure \cite{auer07,cheon07}. The formation of ordered assemblies
starts with the pairing of two peptides, from which larger
$\beta$-sheets develop (Fig.  \ref{fig:11runs}b). As the simulation
progresses, the height of the peak in the size distribution function
associated with single $\beta$-sheets decreases and multi-layer
$\beta$ sheets form, thus revealing the process of protofilament
formation (Fig.  \ref{fig:11runs}c).  This observation complements and
extends the analysis shown in Fig. \ref{fig:Hsheets}, which shows that
the $\beta$ sheets align as they grow in size.

\section*{Discussion}

Although the presence of disordered aggregates might not always be a
prerequisite for amyloid fibril formation, these aggregates do seem to
appear as intermediate states in many cases, and indeed it has been
suggested that in some cases they may serve as initiation sites for
amyloid fibril growth \cite{serio00s,zhu02}.  The simulations that we
present provide molecular details of a sequence-independent mechanism
of formation of amyloid-like structures from the initial disordered
aggregates. This mechanism depends on the interplay between
hydrophobic forces that favor an amorphous collapse and hydrogen
bonding that favor the formation of the ordered cross-$\beta$
structure characteristic of amyloid fibrils.  The $\beta$-sheets that
form within disordered oligomers tend to align into protofilaments,
which then can twist around each other as their lengths increase. In
many protein systems this mechanism will be modulated by the presence
of additional interactions, such as steric repulsions or side chain
hydrogen bonding which are highly sequence specific, but the results
that we present show
that such a mechanism can emerge as a generic feature common
to polypeptide chains.  This phenomenon
thus appears to be an example of the Ostwald step rule in first order
phase transitions \cite{auer04} in which the metastable intermediate
phase from which nucleation takes place is represented by the
disordered compact and highly dynamical oligomeric assemblies that
form prior to the establishment of the ordered cross-$\beta$ amyloid
structure. The general nature of this type of mechanism thus provides
a rationalisation of the observation that oligomeric assemblies appear
to share common structural features, 
including those that enable them to bind to the
same antibodies independently from the sequences of their constituent
peptides and proteins \cite{kayed03}.

In summary, in this work we have investigated the consequences of the
generic hypothesis of amyloid formation \cite{dobson99a} by adopting a
model of protein structure specifically designed to capture the
characteristic of polypeptide chains that are common to all peptides
and proteins \cite{hoang04}.  Our results have provided further
support to the view that the
presence of partially ordered oligomeric assemblies of the type
associated with neurotoxicity constitutes a generic aspect of
the phenomenon of polypeptide aggregation.

\section*{Materials and Methods}

\subsection*{Description of the model}

The tube model
only considers interactions that are common to all polypeptide chains,
and does not include biases towards specific configurations.
In the model~\cite{hoang04} each residue is represented by a
$C_\alpha$ atom.  The atoms are connected into a chain (the protein
backbone) with a fixed distance of $3.8$\AA\ between neighboring
atoms. The lines joining the $C_{\alpha}$ atoms constitute the axes of hard
spherocylinders (cylinders capped by hemispheres) of diameter $4$\AA.
Spherocylinders that do not share a $C_{\alpha}$ atom are not allowed
to interpenetrate. Bond angles are restricted between $82^\circ$ to
$148^\circ$, and bending stiffness is introduced by an energetic
penalty, $e_{\rm S}, >0$ for angles less than $107.15^\circ$; these
are the same criteria used in the original formulation of
the tube model~\cite{hoang04}. Hydrophobicity
enters through a pairwise-additive interaction energy of $e_{\rm HP}$
(positive or negative) between any pair of residues $i$ and $j>i+2$
that approach closer than $7.5$\AA.

The cylindrical symmetry of the tube is broken by the presence of hydrogen bonds.  A hydrogen bond has an energy $e_{\rm
  HB}<0$ and is considered to exist between a pair of residues when
the two normal vectors defined by each $C_\alpha$ atom and its two
neighbors are mutually aligned to within $37^\circ$ and at the same
time each of these vectors lies within $20^\circ$ of the vector
joining the $C_\alpha$ atoms. These geometrical requirements were
deduced from a study of native protein structures \cite{hoang04}.  
There is also
a distance criterion, which is different for local hydrogen bonds
(between residues $i$ and $j=i+3$), and non-local ($j>i+4$) hydrogen
bonds. No more than two hydrogen bonds per residue are permitted, and
the first and last $C_{\alpha}$ atom cannot form inter-chain hydrogen
bonds. Hydrogen bonds may form cooperatively between residues $(i,j)$
and $(i+1,j+1)$, thereby gaining an additional energy of $0.3e_{\rm
  HB}$.  For details of the distance and angle criteria, the reader is
referred to Table 1 of the original article on the tube model~\cite{
hoang04}.

To set the energy scale of the model, the energy of a hydrogen bond is
fixed in all simulations at $e_{\rm HB}=-3kT_o$, where $kT_o$ is a
reference thermal energy and $k$ is Boltzmann's constant.  
This value corresponds approximately the energy associated
with a hydrogen bond ($1.5$kCal/mol at room
temperature~\cite{fersht85}). 
Values of the hydrophobicity and stiffness parameters 
$e_{\rm HP}$ and $e_{\rm S}$ are given in units of $kT_o$ 
and the reduced temperature is
$T^*=T/T_o$.  
In all our simulations we set $e_S=0.9$ and $e_{HP}=-0.15$. 
The ratio of a hydrogen bonding energy
to hydrophobic energy is a parameter that we set to 
$e_{\rm HB}/e_{\rm HP}=20$, which is a value commonly used
in simulations of the aggregation process\cite{nguyen04,pellarin06}.

\subsection*{Simulation techniques}

We performed Monte Carlo simulations in the canonical ensemble using
crankshaft, pivot, reptation, displacement and 
rotation moves\cite{auer07}. To
reduce finite size effects we used a cubic box and applied periodic
boundary conditions. In order to analyze the structure of the
oligomers we used a distance criterion to define a disordered
oligomer, which requires two peptides to have a distance of less
than $5$ \AA. Two peptide chains are considered to form a $\beta$-sheet if
they have more than four inter-chain hydrogen bonds with each other.
To define an angle between different $\beta$-sheets we calculated the
relative orientation between neighboring peptides that constitute the
different $\beta$-sheet. Therefore we require that the centers of mass
of two peptides are separated by less than $10\AA$, which is the
typical inter-sheet distance in both native and most amyloid 
systems\cite{chiti06}.
To extract the angle we calculate the dot product of the end to end
vectors of the peptide molecules. If the average angle between two
$\beta$-strands is less than $20$ degree, we assume that the
respective $\beta$-sheets belong to the same
protofilament.


\begin{thebibliography}{10}

\bibitem{chiti06}
Chiti, F \& Dobson, C.~M.
\newblock (2006) {\em Annu. Rev. Biochem.} {\bf 75}, 333--366.

\bibitem{jahn08}
Jahn, T \& Radford, S.~E.
\newblock (2008) {\em Arch. Biochem. Biophys.} {\bf 469}, 100--117.

\bibitem{makin05}
Makin, O.~S, Atkins, E, Sikorski, P, Johansson, J,  \& Serpell, L.~C.
\newblock (2005) {\em Proc. Natl. Acad. Sci. USA} {\bf 102}, 315--320.

\bibitem{dobson99a}
Dobson, C.~M.
\newblock (1999) {\em Trends Biochem. Sci.} {\bf 24}, 329--332.

\bibitem{selkoe03}
Selkoe, D.~J.
\newblock (2003) {\em Nature} {\bf 426}, 900--904.

\bibitem{stefani03}
Stefani, M \& Dobson, C.~M.
\newblock (2003) {\em J. Mol. Med.} {\bf 81}, 678--699.

\bibitem{harper97a}
Harper, J.~D, Lieber, C.~M,  \& Lansbury, P.~T.
\newblock (1997) {\em Chem. Biol.} {\bf 4}, 951--959.

\bibitem{lambert98}
Lambert, M.~P, Barlow, A.~K, Chromy, B.~A, Edwards, C, Freed, R, Liosatos, M,
  Morgan, T.~E, Rozovsky, I, Trommer, B, Viola, K.~L, Wals, P, Zhang, C, Finch,
  C.~E, Krafft, G.~A,  \& Klein, W.~L.
\newblock (1998) {\em Proc. Natl. Acad. Sci. USA} {\bf 95}, 6448--6453.

\bibitem{serio00}
Serio, T.~R, Cashikar, A.~G, Kowal, A.~S, Sawicki, G.~J, Moslehi, J.~J,
  Serpell, L, Arnsdorf, M.~F,  \& Lindquist, S.~L.
\newblock (2000) {\em Science} {\bf 289}, 1317--1321.

\bibitem{haass07}
Haass, C \& Selkoe, D.~J.
\newblock (2007) {\em Nat. Rev. Mol. Cell. Biol.} {\bf 8}, 101--112.

\bibitem{ma02}
Ma, B \& Nussinov, R.
\newblock (2002) {\em Proc. Natl. Acad. Sci. USA} {\bf 99}, 14126--14131.

\bibitem{hwang04a}
Hwang, W, Zhang, S, Kamm, R,  \& Karplus, M.
\newblock (2004) {\em Proc. Natl. Acad. Sci. USA} {\bf 101}, 12916--12921.

\bibitem{buchete05}
Buchete, N, Tycko, R,  \& Hummer, G.
\newblock (2005) {\em J. Mol. Biol.} {\bf 353}, 804--821.

\bibitem{hills07}
Hills, R.~D \& Brooks, C.~L.
\newblock (2007) {\em J. Mol. Biol.} {\bf 368}, 894--901.

\bibitem{nguyen07}
Nguyen, P.~H, Li, M.~S, Stock, G, Straub, J.~E,  \& Thirumalai, D.
\newblock (2007) {\em Proc. Natl. Acad. Sci. USA} {\bf 104}, 111--116.

\bibitem{cheon07}
Cheon, M, Chang, I, Mohanty, S, Luheshi, L.~M, Dobson, C.~M, Vendruscolo, M,
  \& Favrin, G.
\newblock (2007) {\em PLoS Comp. Biol.} {\bf 3}, 1727--1738.

\bibitem{nguyen04}
Nguyen, H.~D \& Hall, C.~K.
\newblock (2004) {\em Proc. Natl. Acad. Sci. USA} {\bf 101}, 16180--16185.

\bibitem{urbanc04}
Urbanc, B, Cruz, L, Yun, S, Buldyrev, S.~V, Bitan, G, Teplow, D.~B,  \&
  Stanley, H.~E.
\newblock (2004) {\em Proc. Natl. Acad. Sci. USA} {\bf 101}, 17345--17350.

\bibitem{pellarin06}
Pellarin, R \& Caflisch, A.
\newblock (2006) {\em J. Mol. Biol.} {\bf 360}, 882--892.

\bibitem{pellarin07}
Pellarin, R, Guarnera, E \& Caflisch, A.
\newblock (2007) {\em J. Mol. Biol.} {\bf 374}, 917--924.

\bibitem{derreumaux07}
P, D \& N, M.
\newblock (2007) {\em J. Chem. Phys.} {\bf 126}, 025101.

\bibitem{luheshi07}
Luheshi, L.~M, Tartaglia, G.~G, Brorsson, A.~C, Pawar, A.~P, Watson, I.~E,
  Chiti, F, Vendruscolo, M, Lomas, D.~A, Dobson, C.~M,  \& Crowther, D.~C.
\newblock (2007) {\em PLoS Biol.} {\bf 5}, 2493--2500.

\bibitem{hoang04}
Hoang, T.~X, Trovato, A, Seno, F, Banavar, J.~R,  \& Maritan, A.
\newblock (2004) {\em Proc. Natl. Acad. Sci. USA} {\bf 101}, 7960--7964.

\bibitem{hoang06}
Hoang, T.~X, Marsella, L, Trovato, A, Seno, F, Banavar, J.~R,  \& Maritan, A.
\newblock (2006) {\em Proc. Natl. Acad. Sci. USA} {\bf 103}, 6883--6888.

\bibitem{auer07a}
Auer, S, Miller, M, Krivov, S.~V, Dobson, C.~M, Karplus, M,  \& Vendruscolo, M.
\newblock (2007) {\em Phys. Rev. Lett.} {\bf 99}, 178104.

\bibitem{auer07}
Auer, S, Dobson, C.~M,  \& Vendruscolo, M.
\newblock (2007) {\em HFSP J.} {\bf 1}, 137--146.

\bibitem{kammerer04}
Kammerer, R.~A, Kostrewa, D, Zurdo, J, Detken, A, Garcia-Echeverria, C, Green,
  J.~D, Muller, S.~A, Meier, B.~H, Winkler, F.~K, Dobson, C.~M,  \& Stenmetz,
  M.~O.
\newblock (2004) {\em Proc. Natl. Acad. Sci. USA} {\bf 2004}, 4435--4440.

\bibitem{fandrich02}
Fandrich, M \& Dobson, C.~M.
\newblock (2002) {\em EMBO J.} {\bf 21}, 5682--5690.

\bibitem{knowles07s}
Knowles, T. P.~J, Fitzpatrick, A.~W, Meehan, S, Mott, H.~R, Vendruscolo, M,
  Dobson, C.~M,  \& Welland, M.~E.
\newblock (2007) {\em Science} {\bf 318}, 1900--1903.

\bibitem{turner03}
Turner, M.~S, Briehl, R.~W, Ferrone, F.~A,  \& Josephs, R.
\newblock (2003) {\em Phys. Rev. Lett.} {\bf 90}, 128103.

\bibitem{dirix05}
Dirix, C, Meersman, F, MacPhee, C.~E, Dobson, C.~M,  \& Heremans, K.
\newblock (2005) {\em J. Mol. Biol.} {\bf 349}, 903--909.

\bibitem{petty05}
Petty, S.~A \& Decatur, S.~M.
\newblock (2005) {\em Proc. Natl. Acad. Sci. USA} {\bf 102}, 14272--14277.

\bibitem{knowles07}
Knowles, T. P.~J, Shu, W, Devlin, G.~L, Meehan, S, Auer, S, Dobson, C.~M,  \&
  Welland, M.~E.
\newblock (2007) {\em Proc. Natl. Acad. Sci. USA} {\bf 104}, XXX.

\bibitem{serio00s}
{Serio {\em et al.}}, T.~R.
\newblock (2000) {\em Science} {\bf 289}, 1317--1321.

\bibitem{zhu02}
Zhu, M, Souillac, P.~O, Ionescu-Zanetti, C, Carter, S.~A,  \& Fink, A.~L.
\newblock (2002) {\em J. Biol. Chem.} {\bf 277}, 50914--50922.

\bibitem{auer04}
Auer, S \& Frenkel, D.
\newblock (2004) {\em Annu. Rev. Phys. Chem.} {\bf 55}, 333--361.

\bibitem{kayed03}
Kayed, R, Head, E, Thompson, J.~L, McIntire, T.~M, Milton, S.~C, Cotman, C.~W,
  \& Glabe, C.~G.
\newblock (2003) {\em Science} {\bf 300}, 486--489.

\bibitem{fersht85}
Fersht, A.~R, Shi, J.~P, Knill-Jones, J, Lowe, D.~M, Wilkinson, A.~J, low,
  D.~M, Brick, P, Carter, P, Waye, M. M.~Y,  \& Winter, G.
\newblock (1985) {\em Nature} {\bf 314}, 235--238.

\end{thebibliography}

\section*{Acknowledgements} 

This work was supported by the Human Frontier Science Program (SA),
the Research Foundation - Flanders (FWO - Vlaanderen) (FM), the
Leverhulme Trust (SA, CMD and MV), the Wellcome Trust (CMD), and the
Royal Society (MV).


\clearpage

\clearpage

\clearpage

\clearpage

\end{document}